\documentclass{aa}  

\usepackage{natbib}
\bibpunct{(}{)}{;}{a}{}{,} 
\usepackage{graphicx}
\usepackage{hyperref}
\usepackage{txfonts}
\usepackage[normalem]{ulem} 
\begin{document} 

  \title{{A candidate to the long sought optical counterpart to the Rapid Burster in the bulge fossil fragment Liller 1\thanks{Based on observations with the
    NASA/ESA HST, obtained under program GO 15231 (PI: Ferraro). The
    Space Telescope Science Institute is operated by AURA, Inc., under
    NASA contract NAS5-26555.}}}
\titlerunning{An optical counterpart candidate to the Rapid Burster in Liller 1}

   \author{Cristina Pallanca
          \inst{1,2}
          \and
          Francesco R. Ferraro
          \inst{1,2}
          \and
          Barbara Lanzoni
          \inst{1,2}     
          \and 
           Mario Cadelano
          \inst{1,2}
          \and
          Craig O. Heinke
          \inst{3}
          \and
          Maureen van den Berg 
          \inst{4}
          \and
          {Jeroen Homan}
          \inst{5}
          \and
          Chiara Crociati
          \inst{6}
          \and
          Sebastien Guillot
          \inst{7,8}
          }

   \institute{
             Dipartimento di Fisica e Astronomia ``Augusto Righi'', Alma Mater Studiorum Universit\`a di Bologna, via Piero Gobetti 93/2, I-40129 Bologna, Italy\\
              \email{cristina.pallanca3@unibo.it}
    \and
             INAF-Osservatorio di Astrofisica e Scienze dello Spazio di Bologna, Via Piero Gobetti 93/3 I-40129 Bologna, Italy
    \and
            Department of Physics, University of Alberta, Edmonton, AB T6G 2G7, Canada
    \and
            Center for Astrophysics | Harvard \& Smithsonian, 60 Garden Street, Cambridge, MA 02138, USA
    \and 
           Eureka Scientific, Inc., 2452 Delmer Street, Oakland, CA 94602, USA 
    \and
             Institute for Astronomy, University of Edinburgh, Royal Observatory, Blackford Hill, Edinburgh EH9 3HJ, UK
     \and
            IRAP, CNRS,  9 avenue du Colonel Roche, BP 44346, F-31028 Toulouse Cedex 4, France
     \and
          Universit\'e de Toulouse, CNES, UPS-OMP, F-31028 Toulouse, France        
      } 
   \date{Accepted September 7, 2025}

\abstract
{We report on the possible identification of the optical counterpart of the Rapid Burster MXB 1730--335 in the stellar system Liller 1. The identification was performed by taking advantage of a set of images acquired with the Hubble Space Telescope/Advanced Camera for Surveys in the optical band, and with the Gemini South Telescope in the near-infrared. The analysis of these images revealed the presence of a star with a position  possibly compatible with the X-ray and radio band coordinates of the Rapid Burster, and showing significant optical variability. According to its location in the color-magnitude diagram, the candidate companion  appears to belong to the young ($\sim$ 1-2 Gyr old)  super-solar metallicity ([M/H]$=+0.3$) sub-population recently discovered in Liller 1.
We discuss the main characteristics of the candidate counterpart and the Rapid Burster binary system as  derived from the available data, also highlighting the need for further coordinated observations to solidly confirm their association and better clarify their physical properties.}


\maketitle

\section{Introduction}\label{Sec:intro}

The  {\it Rapid Burster} (MXB 1730--335; hereafter RB) was discovered about 50 years ago by \citet{lewin76_discovery} and soon after associated  with Liller 1 \citep{liller77}, a massive ($\sim10^6 M_{\odot}$; \citealp{saracinoLiller}) and highly obscured \citep{valenti,drcLiller1} stellar system in the Galactic bulge  (see below for details).
The RB is an accreting binary system hosting a neutron star (NS), and is expected to share most of the characteristics of low-mass X-ray binaries (LMXBs). However, its properties are really surprising  since it shows a unique behavior in terms of X-ray emission (see \citealp{lewin95} for a review). 
In fact, it is the only known source undergoing both type I and type II X-ray bursts \citep{hoffman78}, and it is also one of the only two sources known so far showing type II X-ray bursts (the other source being the so-called bursting pulsar GRO J1744--28; \citealp{fishman95,kouveliotou96}). 

Briefly, type I bursts are characterized by a spectral softening during burst decay, suggesting a decrease in the effective temperature and thus a cooling of the NS atmosphere. They are thought to be due to thermonuclear flashes of material accreted onto the NS surface  \citep{gallowayTypeIburst}. These characteristics are typical of many LMXBs hosting NSs with  a low magnetic field  (\citealp{masetti02}; see also  \citealp{galloway08,bagnoli13,tzand17}).
Type II bursts, with very short recurrence times \cite[$\sim7$ s to $\sim 1$ hr;][]{sala12},  have been suggested not to be powered by thermonuclear burning, 
 but rather to result primarily from
the release of gravitational energy from the inner accretion disc during spasmodic accretion events \citep{guerriero99, hoffman78, marshallLewin78, spruit93, dangelo10}.

Since its discovery, the RB appeared to be a recurrent transient with outbursts lasting a few weeks, followed by quiescent or "off-state" intervals, which generally last $\sim 6$-8 months \citep{lewin93, guerriero99}. This behavior is  possibly due to some storage mechanism of accretion energy  in the disk around the accreting NS, as described in the disk-instability model \citep[see, e.g.,][]{lasota01}.
At the end of 1999 (MJD $\sim51500$) a sudden change in the  outburst recurrence time and in the X-ray peak intensity was detected \citep{masetti02}: the time between consecutive bursts decreased from $\sim 210$ days down to $\sim 100$ days, and the X-ray peak emission decreased by a factor $\sim 2$. Such a behavior may be connected with  an increase in the  quiescent mass transfer rate from the secondary \citep{masetti02}.

Because of its uniqueness, several multi-wavelength studies of the RB have been performed. For instance, the radio counterpart was identified by \cite{moore00_radioPos} and recently studied by \cite{jetradio}, who found evidence that the radio emission is produced by a jet.
\cite{fox01} searched for type I burst oscillations,  finding a modulation near 307 Hz that is indicative, if confirmed, of a spin period  of 3.25 or 6.5 milliseconds, depending on whether the main burst signal is the fundamental or first harmonic of the spin frequency.

 \cite{kulkarni79} reported the detection of six infrared (IR) bursts of about 30 s each during  2.5 hours of observations. According to the authors, the similarities (rise time, duration, and gradual decay of intensity) of these IR bursts with the type I X-ray bursts detected by \cite{lewin76_discovery} and \cite{hoffman78} 
 suggested an association between the X-ray and the IR sources. They also proposed that the detected IR emission  was unlikely to be blackbody radiation, suggesting that the donor star was still hidden in the highly obscured and crowded population of Liller 1.  However,  \cite{kawara84} reported the absence of IR bursts during a type I X-ray burst and ruled out the connection between type I X-ray bursts and the IR bursts  detected by \cite{kulkarni79}. 

\cite{homer01_chandraPos} also looked for the IR counterpart to the RB using archival ESO/NTT and HST/NICMOS images, confirming that none of the detected stars in the adopted error circle show the properties commonly found   for Roche lobe-filling donor companions, such as unusual IR colors and/or photometric variability \citep[e.g.,][]{comM28H,com6397A}.
A few other works were aimed  at constraining   the physical properties of the NS and/or the binary systems. 
For example, \cite{sala12} presented a detailed study of one of the type I bursts,  with the goal to constrain the NS mass and radius: assuming a distance between 5.8 and 10 kpc, they derived a mass  ${\rm M} = 1.1 \pm 0.3\ {\rm M}_{\odot}$ and a radius  ${\rm R} = 9.6 \pm 1.5$ km.
The investigation of \cite{orbitalPeriod},  based on the comparison of the flux decay to the \cite{kingRitter} model, suggested an orbital period in the range 3.5 - 5.5 hours.

In addition to the unique properties of the RB, recent findings  have  unveiled the peculiar nature of the host stellar system   \citep{lillerBFF,lilche}. In fact, in spite of the initial classification as a globular cluster (GC), recent photometric and spectroscopic studies  have demonstrated that Liller 1  is instead a complex stellar system, with a star formation history characterized by multiple bursts of star formation \citep{lillerBFF,SFHliller, origlia02,chimicaLillerMUSE,chimicaLillerXSH,chimicaLillerNIR}. At least two main sub-populations have been found to coexist in this stellar system: a 12 Gyr-old population at sub-solar metallicity (${\rm[Fe/H]}\sim - 0.4$), and a super-solar (with ${\rm[Fe/H]}=+0.3$) much younger component of just 1-2 Gyr, which is also more centrally segregated \citep{lillerBFF,drcLiller1}. 
The striking chemical similarity between Liller 1 and the bulge field proves a deep connection between the two structures \citep{lilche} and suggests that (similarly to Terzan 5;  \citealp{fer+09,fer+16,origlia+13,terche}) Liller 1 could be a ``bulge fossil fragment'', namely the remnant of one of the building blocks that contributed to the formation of our spheroid at the epoch of the Milky Way  assembly. 
This new emerging scenario further enhances the uniqueness of the RB, which turns out to be peculiar not only in its own properties, but also with respect to its hosting environment.

In this paper, we report on the possible identification of the optical counterpart to the RB obtained from the analysis of a set of optical and near-IR high-resolution images secured in the core of Liller 1.
The used  datasets and catalogs are reported  in Section \ref{dataset}, while  Section \ref{companion} presents the properties of the star proposed as the RB counterpart.  Finally, Section \ref{discussionMain} summarizes the key findings.

\begin{table*}
\centering  
\caption{ Most recent X-ray and radio band positions of the RB, and coordinates of its candidate optical counterpart.}
\label{Tab:positions}
\begin{tabular}{ccllr} 
\hline \hline
	R.A. (J2000)& Dec (J2000)	& spectral band/instrument &   Reference \\ 
     $\rm[h\ m\ s]$ & $\rm[^\circ\ '\ '']$ & & \\
 	 \hline
   $17\ 33\ 24.61 \pm 0.03$ & $-33\ 23\ 19.9 \pm0.4$ &  X-ray/Chandra &\cite{homer01_chandraPos} \\ 
   $17\ 33\ 24.61 \pm 0.01$ & $-33\ 23\ 20.1 \pm0.6$ &  Radio/VLA & \cite{jetradio} \\
   $17\ 33\ 24.66 \pm 0.01$ & $-33\ 23\ 19.84 \pm0.15$   & Optical/HST & This work
\end{tabular}
\end{table*}

\section{Observations and data analysis}
\label{dataset}
 Table \ref{Tab:positions} reports the absolute coordinates of the RB as obtained from the most recent X-ray and radio band observations. They
agree  on localizing the RB in the most crowded region of Liller 1, at only $2.8\arcsec$ from its gravitational center \citep{saracinoLiller}. 
This implies that high angular resolution images are mandatory to properly resolve individual stars and search for the optical counterpart to the RB, and 
we therefore took advantage of the multi-wavelength (optical and IR) dataset presented in \cite{drcLiller1} and already used in \cite{lillerBFF} and \cite{SFHliller}.
The optical dataset  was acquired with the HST/ACS-WFC under program GO 15231 (PI: F.R.  Ferraro), and it consists  of 6 images of about 1330 s each in the F606W filter, and 6 images of 800 s in the F814W filter. 
The IR dataset consists  of $6 \times 30$ s images in the Ks filter acquired with the Gemini South Adaptive Optics Imager (GSAOI) assisted by the Gemini Multi-Conjugate Adaptive Optics System (GeMS), mounted at the Gemini South Telescope (Program ID: GS- 2013-Q-23; PI: D. Geisler).

Due to  high crowding, the photometric analysis has   necessarily been performed with the Point Spread Function (PSF) fitting method, by adopting a  synergistic approach aimed at  extracting the most information from different wavelengths. In the following, we briefly summarize the main steps of the photometric analysis \citep[see][for more details]{drcLiller1}.
First of all, we modeled the PSF on a sample of isolated, bright stars in each secured frame.  The obtained PSF model was then applied to the master-list of stellar objects identified in the images. 
The latter has been built by  compiling all the sources detected in at least one of the two datasets (i.e., optical and IR), with the aim to enhance 
the completeness of different spectral types. 

The instrumental magnitudes have been calibrated to the VegaMag system and corrected for differential reddening as described in \cite{drcLiller1}. To improve the astrometric precision of the photometric catalog, we also  aligned the original astrometry (which was referred to Gaia DR2; \citealp{gaiaDR2}) to the most recent release Gaia DR3 \citep{gaiaDR3}. Individual magnitudes were finally homogenized to the first reference image of each frame by using the DAOPHOT routine DAOMASTER \citep{daophot,daophotAllframe}. In addition to the original photometry, we also computed the magnitudes and the corresponding MJD of single frames, with the aim to perform a detailed variability analysis of any promising candidate counterpart to the RB.

\section{The candidate optical counterpart to the RB}\label{companion}

\begin{figure}[h]
\includegraphics[width=\hsize]{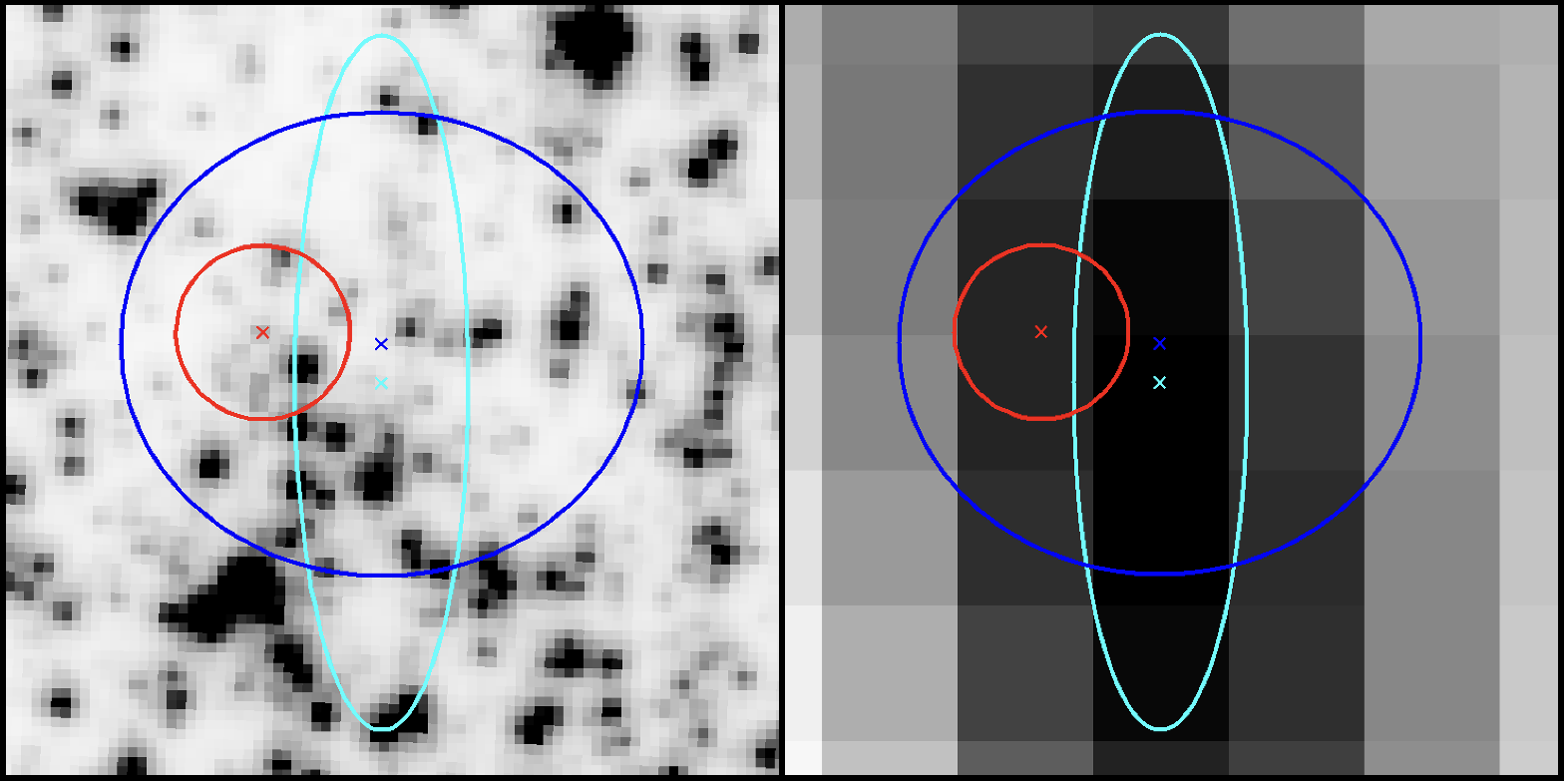}
\centering
\caption{HST/ACS (left panel) and VLA   (right panel; \href{https://zenodo.org/records/12806382}{https://zenodo.org/records/12806382})  images of the $4\arcsec \times 4 \arcsec$ region centered on the X-ray position of the RB quoted in Table \ref{Tab:positions}. The crosses mark the coordinates listed in the table, while the ellipses have axes equal to 3 times the quoted uncertainties, with red, blue and cyan colors corresponding, respectively, to the values obtained for the candidate optical counterpart (this work), and the X-ray and radio band  positions of the RB \citep[][respectively]{homer01_chandraPos, jetradio}. North is up, East is to the left.}
\label{Fig:map}
\end{figure}

\begin{figure}[h]
\includegraphics[width=\hsize]{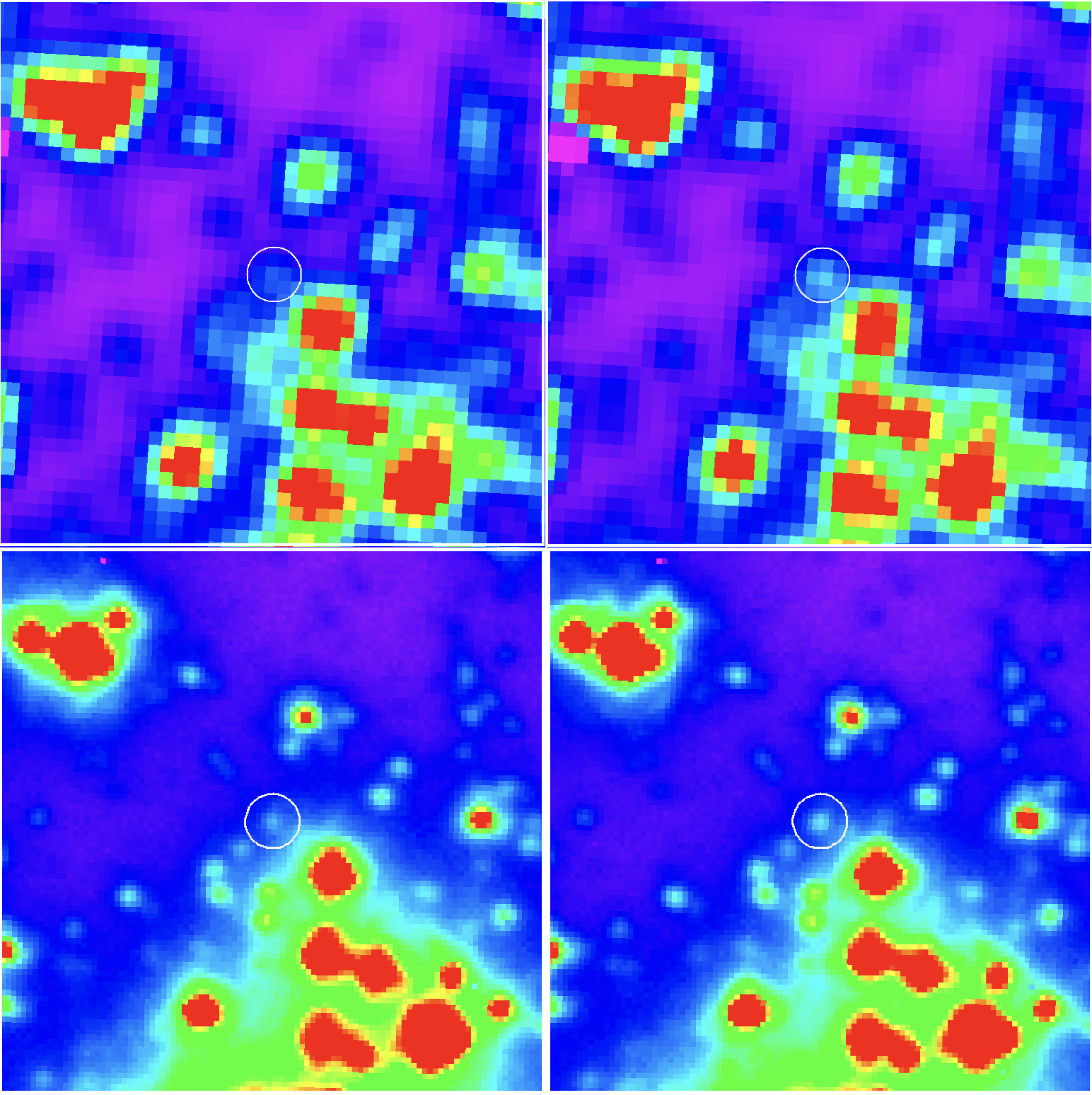}
\centering
\caption{ Images of the $2\arcsec \times 2\arcsec$ region surrounding the candidate optical counterpart to the RB (white circle) acquired  at different times. 
The top and bottom panels show HST/ACS images in the F814W filter, and Ks-band GeMS images, respectively. For visualization purposes, a 2 pixel smoothing has been applied to the F814W images. The images in the right panels were acquired approximately 1.5/2 hours after those in the left panels. The corresponding changes in magnitudes are about $\sim 0.4$ mags for the Ks-band (bottom row) and $\sim0.5$ mags for the F814W filter (top row).}
\label{Fig:mapVAR}
\end{figure}

\begin{figure}[h]
\includegraphics[width=\hsize]{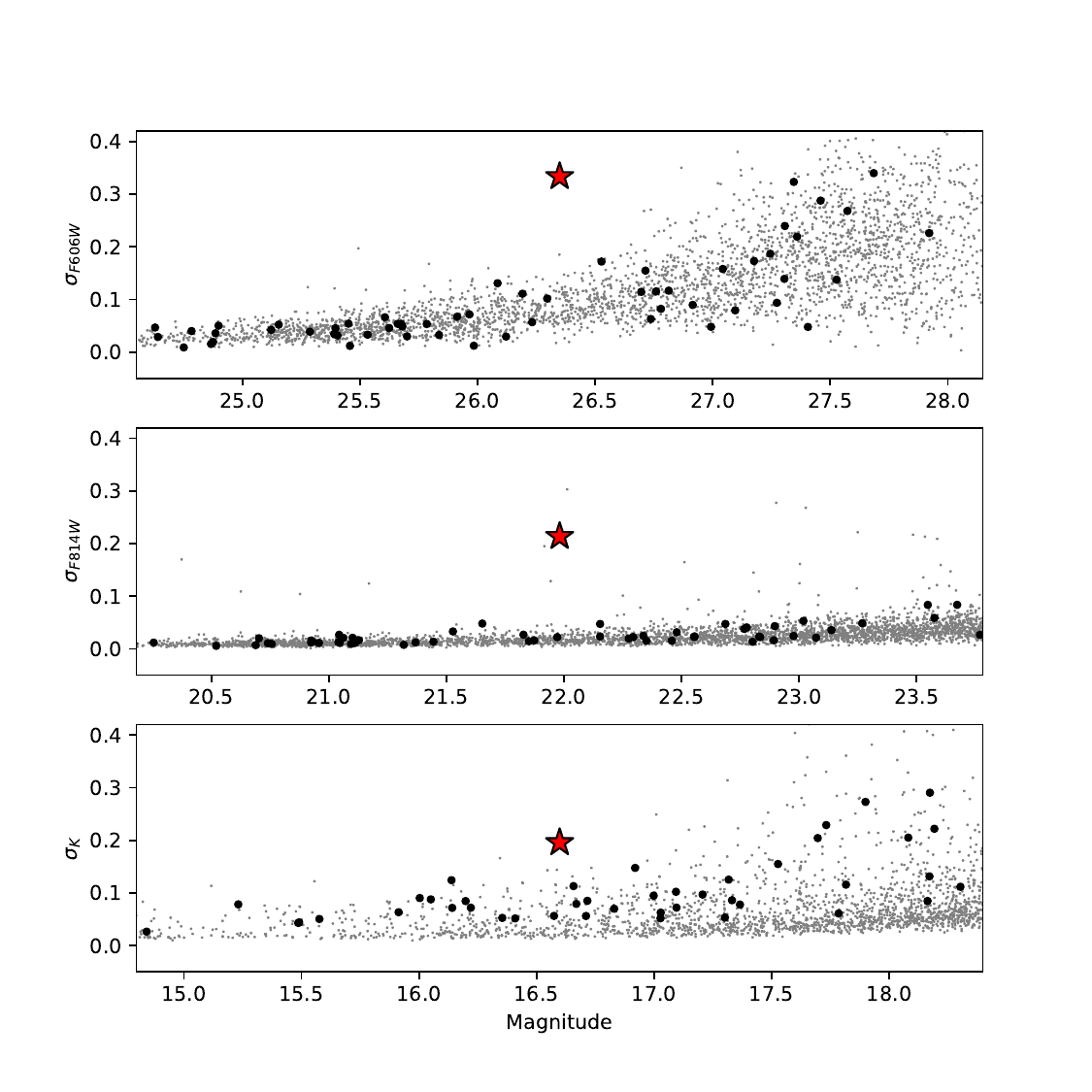}
\centering
\caption{ Magnitude scatter measured for all the stars located  within $20\arcsec$ (gray dots) from the RB X-ray position, as a function of their magnitude. The black circles highlight all the sources detected within a distance of just $1.5\arcsec$ from the RB, with the large red star marking the candidate optical counterpart. The top, middle and bottom panels refer to the F606W, F814W, and Ks filters, respectively.}
\label{Fig:varIndex}
\end{figure}

\begin{figure}[h]
\includegraphics[width=\hsize]{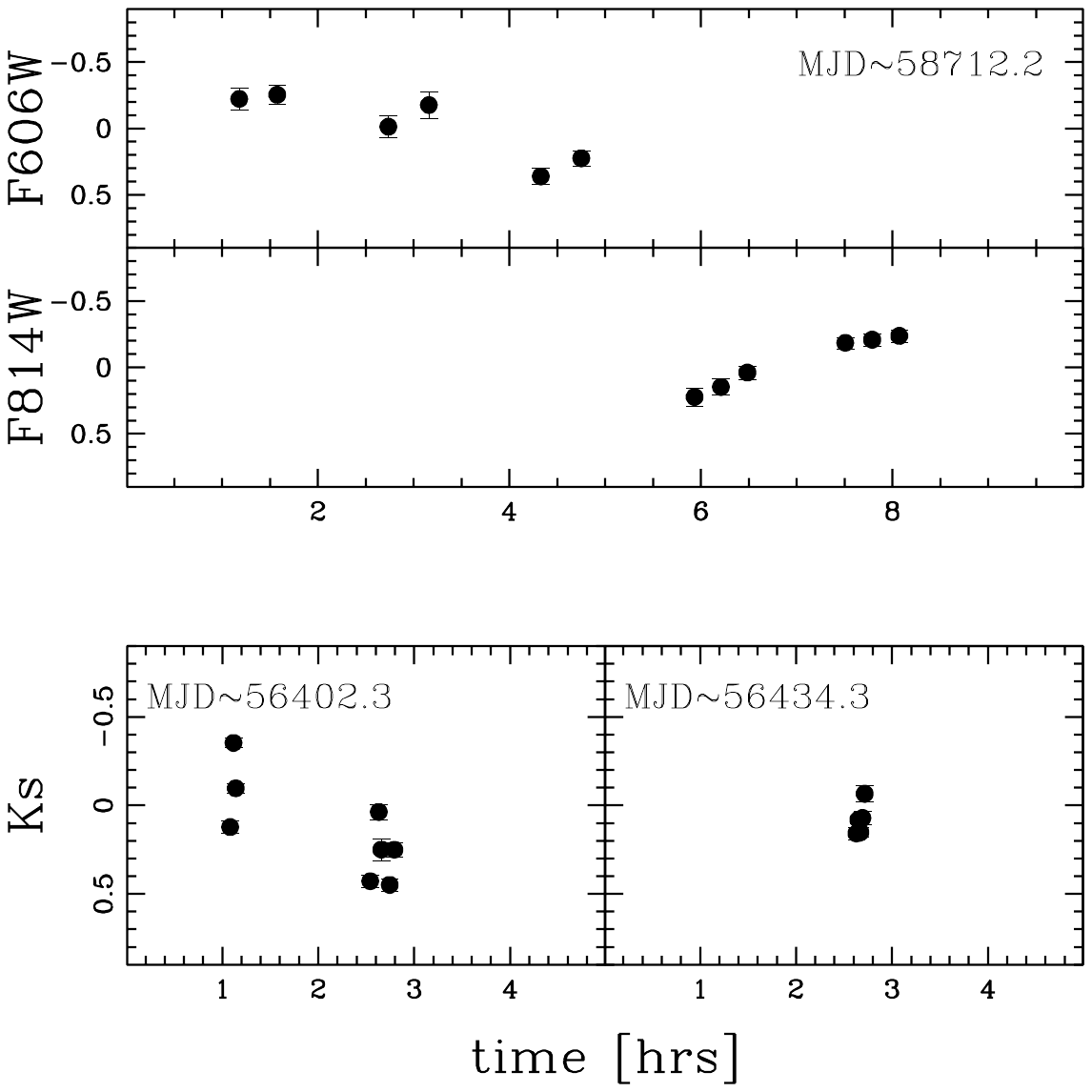}
\centering\caption{Magnitude differences with respect to the mean value in the F606W, F814W, and Ks filters (top, middle, and bottom panels, respectively), as a function of time. The latter is calculated in hours with respect to the reference MJD marked in each panel. Error bars for the magnitudes are displayed and are comparable in size to the plotted symbols} 
\label{Fig:lcPROPOSAL}
\end{figure}

The identification of optical counterparts to exotic objects (like X-ray binaries, millisecond pulsars, cataclysmic variables, etc.) in high-density environments is  a 
complex task because a large number of stars is typically detected within the positional error box of the investigated source. 
 Moreover, the radio, X-ray and optical astrometry of the same source can suffer from some $0.1\arcsec$ misalignment (see, e.g., \citealp{edmonds01, edmonds03, albrow01, huang10}).
Thus, beside the reasonable positional coincidence   (within the errors) of the optical position with the  radio  and/or X-ray  positions, one of the 
best features for the identification of the optical counterpart to exotic objects is the detection of a luminosity modulation 
connected with the orbital motion and/or the rotation of the source \citep[see][]{comM28H,Ha6397,com6397A,comM71A}. 
Moreover, the optical counterparts to interacting/transient binaries typically are highly perturbed/bloated/distorted/irradiated objects, and their magnitudes arise from the blending of the two stellar components and/or with an accretion disk. For these reasons, they are also expected to appear in anomalous positions in the color-magnitude diagram (CMD) with respect to the canonical sequences traced by ``normal'' cluster stars (see example in \citealt{comM28H,comM28I,EXOter5,rivera1547Tuc,AMSP6440,lugger23M4,ettorre362}).

The left panel of Figure \ref{Fig:map} shows the $4\arcsec \times 4\arcsec$ region of a F814W HST/ACS image centered on the X-ray position of the RB reported in Table  \ref{Tab:positions}. The blue and cyan ellipses have axes equal to 3 times the R.A. and Dec uncertainties quoted in the table. 
 As can be  seen, several stars are detected within these error ellipses. Hence, to identify the optical counterpart to the NS, we searched for evidence of photometric variability.

Indeed, a visual inspection of these images already provided us with 
hints of variability for one source.
This can be appreciated in Figure \ref{Fig:mapVAR}, where the star marked with a white circle appears to be brighter in the right-hand panels than in the left-hand ones, both in the F814W optical HST/ACS images (upper panels), and in the Ks-band GEMINI exposures (bottom panels). 
The location of this  variable source with respect to the X-ray and radio-band positions of the RB is marked by the red cross and circle in Figure \ref{Fig:map}.
To assess the significance of its variability, we used the external error provided by DAOPHOT, which 
 quantifies the magnitude scatter among different exposures. For variable objects, this parameter is expected to be larger than that of non-variable stars of similar mean magnitude. Figure \ref{Fig:varIndex} shows the 
values of the magnitude scatter calculated for all the stars observed within $20 \arcsec$ from the RB (gray dots), as a function of their mean magnitude, in the F606W, F814W and Ks bands. Clearly, among all the stars located within $1.5\arcsec$ from the RB position (black circles), the detected source (red star in the figure) is the only one showing a magnitude scatter significantly larger than that of the objects with similar magnitude. 

In Figure \ref{Fig:lcPROPOSAL} we plot the  magnitude differences measured in the three investigated filters as a function of time. The observations secured through the F606W and the F814W filters show, respectively, a decrease and an increase of luminosity on a timescale compatible with the orbital period (3.5-5.5 hr) proposed by \cite{orbitalPeriod}. 

The K-band magnitudes show substantial scatter on short timescales, which is not physically plausible for emission from the donor star (which should dominate at these wavelengths). 
Although we do not have sufficient information to clearly explain this behavior, we can speculate that it may be due to the presence of a dominant non-thermal component \citep[such as a jet;][]{kulkarni79}, or thermal reprocessing by an accretion disk \citep{vanParadijsMcClintock}, which is expected to manifest as strong flickering. 
In the latter case, the system would be in outburst and the observed optical/NIR flux would likely be dominated by the accretion disk, thus giving us little information on the donor.
In addition, we notice that if the donor is significantly heavier than the accretor (as the counterpart mass may suggest; see Section \ref{discussionMain}), mass transfer would be unstable. In any case, it must be acknowledged that the K-band variability is less distinctive, compared to other stars (see Figure \ref{Fig:varIndex}, suggesting that the short-term K-band variations may not be real. Unfortunately, given the short time sampling of the available dataset and the relatively long exposure times, no firm conclusions on the characteristics of the light curve can be reached. Future simultaneous optical/near-IR observations
\citep{russel06} are required to fully dissect this binary system but there is no doubt that this star presents the typical signature of a perturbed object.

\begin{figure}[h]
\includegraphics[width=\hsize]{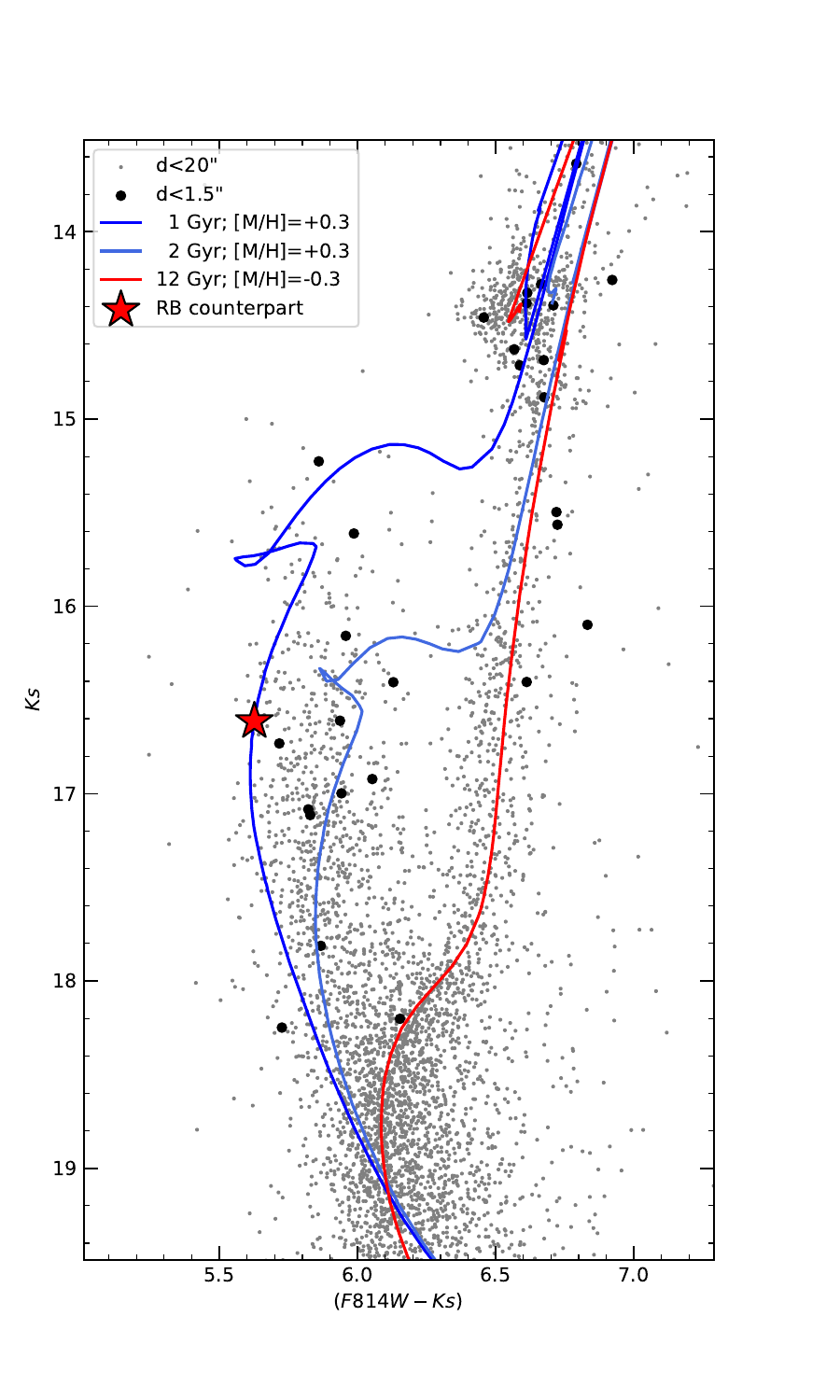}
\centering
\caption{Differential reddening corrected and proper motion selected CMD of all the stars detected within $20 \arcsec$ from the RB X-ray position (gray dots). The black circles and the large red star highlight, respectively, the sources located within a distance of $1.5\arcsec$ and the candidate optical counterpart to the RB. The red and blue lines  (see also legend) are, respectively, a 12 Gyr old isochrone \citep{Bressan+12} with [M/H]=$-0.3$ (well reproducing the old sub-population of Liller 1), and  two  1-2 Gyr old isochrones with [M/H]=$+0.3$ (tracing the young and super-solar sub-population). }
\label{Fig:cmd}
\end{figure}

\section{Discussion and conclusions }
\label{discussionMain}
The  variable source detected in the HST and Gemini images is located at ${\rm RA} = 17^{\rm h} 33^{\rm m} 24.66^{\rm s}$, ${\rm Dec}= -33^{\circ}23\arcmin19.84\arcsec$ (J2000.0) in the ICRS GAIA DR3 astrometric system, with an uncertainty of $0.15\arcsec$ in both coordinates (see the red circle in Figure 1). 
Its declination is consistent with the values estimated in the X-ray and radio bands, while a larger offset is detected along the right ascension direction (see Table \ref{Tab:positions}). The agreement with the X-ray R.A. value is still within $\sim 1.5 \sigma$.  We also notice that the X-ray coordinates quoted in Table \ref{Tab:positions} have been shifted by $0.56\arcsec$ in R.A. and $0.08\arcsec$ in Dec with respect to the nominal Chandra coordinates 
(see the discussion in  Section 2.2 of \citealt{homer01_chandraPos}),  thus pointing to a possible additional source of uncertainty in the X-ray position.
 Instead, the right ascension values in the optical and radio-bands are formally only consistent at 3$\sigma$, with the VLA uncertainty quoted in Table \ref{Tab:positions} corresponding to 10\% of the radio beam size when using all baselines (see \citealp{jetradio}). 
We notice, however, that the coordinates provided in Section 2.1 of \citet{jetradio} for a radio source used to check for the RB variability slightly differ from those quoted in their Table 2. 
 Moreover, the emission shown in the observed radio map  (see the right-hand panel of Figure \ref{Fig:map})
 looks well compatible with the position of the variable star detected in this study.  
Taking into account all the uncertainties, we therefore conclude that the position of the detected optical source is consistent, at worse within 1.5 $\sigma$, with the X-ray coordinates of the RB and, at worse within 3 $\sigma$, with its radio-band position.

This source is the only one showing significant photometric variability among all the stars detected within a distance of $\sim 1.5\arcsec$ from the RB (Figure \ref{Fig:varIndex}). It shows light variations in all the investigated filters, with  different properties at different wavelengths. In particular, in the optical band it presents modulations on a timescale of hours, which is compatible with the orbital period estimated by \cite{orbitalPeriod}  as well as with larger values (up to one day), and it is consistent with typical orbital motion of compact binaries, suggesting heating and/or tidal distortions  resulting in ellipsoidal variability  \citep{kennedy18,ellipsoidal}. At larger wavelength (Ks band), the variability appears to occur on shorter timescales, and it is worthy of further investigation to 
 verify the reality of the short-term variations, and if real, understand its timescale and nature
\citep[due, e.g., to some jet-related emission processes;][]{baglio14}.  Based on the reasonable positional coincidence and the detected variability, we conclude that this star is a promising candidate for being the RB counterpart.
Unfortunately, a firm and unambiguous association with the RB is prevented by the still unknown orbital period of the binary system  and the incomplete sampling of the optical light curve. In this context, new observations aimed at better characterizing the light curve at different wavelengths simultaneously would be crucial to both determine the variability period and better understand the accretion processes driving this unique system.
Even more interesting would be the detection, at the position of the candidate counterpart  quoted here (see Table \ref{Tab:positions}), of radio pulses with the spin frequency proposed by \cite{fox01}  and compatible with an orbital motion consistent with the magnitude modulations.

To search for possible connections between the optical variability of the candidate counterpart and the X-ray variability of the RB, we compared the MJD of observations here discussed (see Figure\ref{Fig:lcPROPOSAL}) with the 
 dates of peak flux observed by MAXI
reported in \cite{heinkeListBurst}. 
Unfortunately, each of the three observations are in between the known outbursts seen with MAXI, at least one month apart from the nearest outburst peak. Even if the MAXI data  are not very constraining on the RB flux (due largely to absorption, and partly to crowding;  see \citealt{negoro16MAXI} for all details about the MAXI data analysis), 
 according to
the MAXI datapoints acquired in the closest MJD epochs to the optical/IR observations, 
 we roughly estimated upper limits on the corresponding
X-ray fluxes, finding: 
 $L_{\rm x}<3\times10^{36}$ erg s$^{-1}$ ($\rm MJD\sim 56402$), $L_{\rm x}<10^{37}$ erg s$^{-1}$ ($\rm MJD \sim 56434$), and 
$L_{\rm x}<5\times10^{36}$ erg s$^{-1}$ ($\rm MJD\sim 58712$). 
Unfortunately, this brings no conclusive results and calls  for future (almost) simultaneous X-ray and optical/IR observations, which are essential both for  investigating the association of the proposed counterpart with the RB, and for gaining a better understanding of the physics governing the evolution of this binary system.

Although the variability sampling is admittedly non-optimal, we estimate that the candidate counterpart to the RB has mean magnitudes $m_{\rm F606W}=26.3$, $m_{\rm F814W}=22.0$, and $\rm K=16.6$. Figure \ref{Fig:cmd} shows its position (marked with a large red star) in the differential reddening corrected and proper motion selected CMD.
As can be seen, it is fully compatible with the blue plume of Liller 1, corresponding to the main sequence of the very young and super-solar 
sub-component recently discovered in this system \citep{lillerBFF}. 
Its location along the main sequence of a 1 Gyr old isochrone \citep{Bressan+12} with global metallicity [M/H]$=+0.3$ (blue line in Figure \ref{Fig:cmd}) suggests a mass of $\sim 2\ M_\odot$, a surface temperature of $\sim7200$ K, and a radius of $2.8\ R_\odot$ for this star.

Assuming these parameters and using the orbital period/radius relation of \cite{FrankKingRaine},  a binary system with an orbital period of about 0.5-2 days would be required. This is larger than the orbital period suggested by \citet{orbitalPeriod}. However, we notice that, at odds with the results discussed in \cite{orbitalPeriod}, the overall outburst behavior of the RB \citep{periodGRS1747} is fairly similar to that of Aquila X--1 and GRS 1747--312, which have orbital periods of 19 hours \citep{periodAq1} and 12 hours \citep{periodGRS1747}, respectively. This suggests that a larger period seems to be plausible. 

On the other hand, there is also the possibility  that the mass and radius derived above are not fully reliable because the observed CMD position is altered by distortion/interaction effects with the accreting NS, or the presence of additional components of the binary system (such as an accretion disk), in agreement with what observed in other interacting binaries (e.g., \citealt{com6397A, edmonds02,bassa04,comM28H,comM28I,beccari14,EXOter5,kumawat362}). 
Curiously, the position of this star in the CMD is very similar to that of the X-ray burster EXO~1745--248 in Terzan 5 during the outburst phase \citep{EXOter5}. However, according to MAXI all-sky X-ray monitoring, in all the epochs sampled by the observations discussed in this paper, the RB was either in quiescence, or in the tail of an outburst. Hence, the observed optical emission should largely come from the RB donor star (in place of accretion or jets). This would support the hypothesis that the donor star of the RB system is a metal-rich object formed $\sim 1$ Gyr ago during the most recent star formation burst occurred in Liller 1 \citep{SFHliller}.
This could have important consequences for the characterization of the binary system and the accreting NS, possibly suggesting that it also formed very recently instead of being the remnant of a supernova explosion occurred at the main epoch of star formation, $\sim 12$ Gyr ago (see also the discussion by \citealt{Patruno12_IGRter5} for the case of IGR J17480--2446 in Terzan 5). Further modeling efforts may explore whether the physical properties of such a young system are compatible with the unusual magnetic field characteristics of the RB, often associated with type-II burst mechanisms.

The distance of Liller 1 is one of the key ingredients used by  \citet{sala12} to constrain the physical properties of the NS hosted in the RB binary system from the observed photospheric radius expansion.
These authors explored a range of distances between 5.8 and 10 kpc, obtaining a mass $M = 1.1 \pm 0.3M_\odot$ and a radius  $R = 9.6 \pm 1.5$ km. 
From the probability density distribution plotted in the NS mass-radius diagram (see their Figure 7), the most recent distance estimate \citep[$\sim 8.5$ kpc;][]{lillerBFF} would favor the upper limits to the quoted mass and radius, consistently with recently derived physical properties of millisecond pulsars via X-ray pulse-profile modeling
\citep[e.g.,][for the two most recent measurements]{choudhury24,mauviard25}.  However, it is important to emphasize that other methods are commonly used to estimate the NS properties \citep[e.g.,][]{NSmassaRaggio,marino18}, and further studies are necessary for the RB case.

We emphasize that only a few X-ray sources (see \citealt{homer01_chandraPos}) and no millisecond pulsars have been detected so far in Liller 1. 
This is somehow 
peculiar since this stellar system is the only one so far that, together with Terzan 5 (see \citealt{fer+16,terche}), shows the presence of multi-iron \citep{lilche} and multi-age stellar sub-populations \citep{lillerBFF}, and it is considered a bulge fossil fragment. 
The chemical evolutionary model recently developed for Terzan 5 \citep{Romano+23} combined with the reconstructed star formation histories of both systems \citep{SFHTer5,SFHliller}  have demonstrated that their observed chemical patterns can be properly reproduced by assuming a high star formation rate and several $10^4$ explosions of core-collapse supernovae, producing a large number of NSs. The large mass of the  stellar system progenitor (estimated to be a few  \citealp[ $10^7 M_{\odot}$;][]{Romano+23}) allowed the retention of such a large population of remnants, and the high collision rate \citep{VerbuntHut87,Lanzoni+10}  providing the ideal environment for the formation of binaries with (at least) one compact object. Indeed, this scenario naturally explains the striking populations of X-ray sources  \citep{Ter5Xray}  and millisecond pulsars \citep{MSPter5,MSPter52} detected in Terzan 5, and 
in principle, a similar framework would be expected also for Liller 1.

However,  this apparent discrepancy may be due to several observing challenges. In fact, Liller 1  is moderately distant, quite extincted \citep{saracinoLiller,drcLiller1} and, most importantly, it has a  bright transient  X-ray source that makes it difficult to study the other faint X-ray sources in the system. It is also worth mentioning that \citep{fruchterGoss00} reported a steep-spectrum radio emission from Liller 1, which, however, was not found by \citep{mccarver24}, with observations acquired at higher angular resolution. This suggests that the observed emission may originate from several pulsars, which remain still undetected because of strong scattering and likely large (but still unconstrained) dispersion measure.

Understanding the physical phenomena characterizing the RB could be fundamental in the context of clarifying the properties and evolutionary paths of other exotica populating GCs, such as the fast radio burst (FRB) detected in an extragalactic GC by \citet{zhangFRBM81}, 
a binary millisecond pulsar with a NS/black hole companion
\citep{1851E}, or, more  generally, binaries containing compact objects and their potentiality as GW emitters. For example, \cite{FRBinGC} suggest that FRBs in GCs could be generated from a newly born NS (formed through the accretion induced collapse scenario; see \citealp{tauris13AIC}) that could still reside in a  binary system. In this case, a fraction of GC FRB sources should have binary companions, producing interesting observational consequences, such as periodicity in burst repetition \citep[e.g.,][]{lyutikov20}, persistent X-ray emission (from possible subsequent accretion through Roche lobe overflow; e.g., \citealp{tauris13AIC}), or time-dependent dispersion measure/rotation measure. Hence, the investigation of a relative nearby binary system hosting a unique NS, such as the  RB, may be helpful also to interpret some peculiar properties of more elusive extragalactic sources.

\begin{acknowledgements}
We thank the referee for the very helpful suggestions.
This work is part of the project Cosmic-Lab at the Physics and Astronomy Department ``A. Righi'' of the Bologna University (http://www.cosmic-lab.eu/ Cosmic-Lab/Home.html). COH is supported by NSERC Discovery Grant RGPIN-2023-04264.

\end{acknowledgements}


\bibliographystyle{aa} 
\bibliography{biblio.bib} 
\end{document}